\documentclass[lettersize,journal]{IEEEtran}
\usepackage{amsmath,amsfonts}
\usepackage{algorithmic}
\usepackage{algorithm}
\usepackage{array}
\usepackage[caption=false,font=normalsize,labelfont=sf,textfont=sf]{subfig}
\usepackage{textcomp}
\usepackage{stfloats}
\usepackage{url}
\usepackage{verbatim}
\usepackage{graphicx}
\usepackage{cite}
\usepackage[colorlinks=true, linkcolor=red, citecolor=green]{hyperref}
\usepackage{multirow}
\usepackage{booktabs} 
\usepackage[table]{xcolor}         
\usepackage{stackengine}
\usepackage{wrapfig}
\usepackage{xspace}

\newcommand{\onedot}{.\@\xspace}

\newcommand{\ie}{\emph{i.e}\onedot}

\newcommand{\etal}{\emph{et al}\onedot}

\definecolor{myblue}{RGB}{66,133,244}
\definecolor{mygreen}{RGB}{51,168,83}
\definecolor{myyellow}{RGB}{251,188,3}
\definecolor{myred}{RGB}{234,67,53}
\definecolor{mygrey}{RGB}{95,99,104}

\begin{document}

\title{CMamba: Learned Image Compression with State Space Models}

\author{Zhuojie Wu, Heming Du, Shuyun Wang, Ming Lu, Haiyang Sun, Yandong Guo, Xin Yu
\thanks{Zhuojie Wu, Heming Du, Shuyun Wang, and Xin Yu are with the School of Electrical Engineering and Computer Science, University of Queensland, Brisbane 4067, Australia (e-mail:
zhuojie.wu@uq.edu.au; heming.du@uq.edu.au; shuyun.wang@uq.edu.au; xin.yu@uq.edu.au). \textit{(Corresponding author: Xin Yu.)}

Ming Lu is with Intel Lab China, Beijing 100876, China (e-mail: lu199192@gmail.com).

Haiyang Sun is with LiAuto, Shanghai 201805, China (e-mail: sunsea48@gmail.com).

Yandong Guo is with AI$^2$ Robotics, Shenzhen 518055, China (e-mail: yandong.guo@live.com).

This work has been submitted to the IEEE for possible publication. 
Copyright may be transferred without notice, after which this version may no longer be accessible.
}
}

\markboth{Journal of \LaTeX\ Class Files,~Vol.~14, No.~8, August~2021}%
{Shell \MakeLowercase{\textit{et al.}}: A Sample Article Using IEEEtran.cls for IEEE Journals}

\IEEEpubid{0000--0000/00\$00.00~\copyright~2021 IEEE}

\maketitle

\begin{abstract}

Learned Image Compression (LIC) has explored various architectures, such as Convolutional Neural Networks (CNNs) and transformers, in modeling image content distributions in order to achieve compression effectiveness.
However, achieving high rate-distortion performance while maintaining low computational complexity (\ie, parameters, FLOPs, and latency) remains challenging.
In this paper, we propose a hybrid Convolution and State Space Models (SSMs) based image compression framework, termed \textit{CMamba}, to achieve superior rate-distortion performance with low computational complexity.
Specifically, CMamba introduces two key components: a Content-Adaptive SSM (CA-SSM) module and a Context-Aware Entropy (CAE) module.
First, we observed that SSMs excel in modeling overall content but tend to lose high-frequency details. 
In contrast, CNNs are proficient at capturing local details.
Motivated by this, we propose the CA-SSM module that can dynamically fuse global content extracted by SSM blocks and local details captured by CNN blocks in both encoding and decoding stages. 
As a result, important image content is well preserved during compression.
Second, our proposed CAE module is designed to reduce spatial and channel redundancies in latent representations after encoding.
Specifically, our CAE leverages SSMs to parameterize the spatial content in latent representations. 
Benefiting from SSMs, CAE significantly improves spatial compression efficiency while reducing spatial content redundancies. 
Moreover, along the channel dimension, CAE reduces inter-channel redundancies of latent representations via an autoregressive manner, which can fully exploit prior knowledge from previous channels without sacrificing efficiency.
Experimental results demonstrate that CMamba achieves superior rate-distortion performance, outperforming VVC by 14.95\%, 18.83\%, and 13.89\% in BD-Rate on Kodak, Tecnick, and CLIC datasets, respectively.
Compared to the previous best LIC method, CMamba reduces parameters by 51.8\%, FLOPs by 28.1\%, and decoding time by 71.4\% on the Kodak dataset.

\end{abstract}
\section{Introduction}

Image compression is a vital technology in multimedia applications, allowing for efficient storage and transmission of digital images. 
With the rise of social media, a large number of images are created by users and transmitted over the internet every second. 
Advanced compression methods are constantly sought to achieve superior rate-distortion performance while maintaining efficiency.
Classical lossy image compression standards, such as JPEG~\cite{wallace1991jpeg}, BPG~\cite{bellard2018bpg}, and VVC~\cite{bross2020versatile}, achieve commendable rate-distortion performance via handcrafted rules. 
With the advances in deep learning, Learned Image Compression (LIC) methods~\cite{balle2017end, song2021variable, cui2021asymmetric, ma2022end, ali2023towards, theis2022lossy, mentzer2018conditional, li2020efficient, son2021enhanced, dardouri2021dynamic} make promising progress and present better rate-distortion performance by exploiting various Convolutional Neural Networks (CNNs) and transformer architectures.

In general, LIC follows a three-stage paradigm: \textbf{nonlinear transformation}, \textbf{quantization}, and \textbf{entropy coding}.
The nonlinear transformation consists of an analysis transform and a synthesis transform.
The analysis transform maps an image from the pixel space to a compact latent space. 
The synthesis transform is an approximate inverse function that maps latent representations back to pixels.
Quantization rounds latent representations to discrete values, and entropy coding encodes them into bitstreams.
In particular, LIC faces two critical challenges: 
(1) how to design an effective yet efficient nonlinear transformation that yields a compact latent representation in the analysis transform and recovers a high-fidelity image in the synthesis transform, and
(2) how to achieve efficient entropy coding for highly compressed bitstreams.

Many studies have sought to address the aforementioned challenges~\cite{zhou2019end, liu2023learned, he2021checkerboard, minnen2020channel}.
As for the first challenge, CNNs based models often struggle to capture global content, causing redundancy in latent representations~\cite{zhou2019end, cheng2020learned}. 
To address this problem, several works leverage transformers for image compression due to their powerful long-range modeling capabilities~\cite{kenton2019bert, dosovitskiy2020image, liu2021swin, zhu2022transformer, zou2022devil, li2024frequency,chen2021end,liu2023learned}. 
However, the quadratic complexity of self-attention incurs high computational cost, thus restricting efficient compression.
As for the second challenge, autoregressive models and transformers are two popular options in exploiting spatial or channel correlations~\cite{minnen2018joint, qian2021entroformer, he2021checkerboard, minnen2020channel, liu2023learned, li2024frequency, jiang2023mlic, koyuncu2022contextformer}.
Since the spatial dimension is often quite large, modeling the spatial dependency in an autoregressive manner will lead to high latency~\cite{minnen2018joint, qian2021entroformer}.
Moreover, existing channel-wise autoregressive models can only remove inter-channel redundancy~\cite{minnen2020channel, zou2022devil}. 
Thus, the spatial redundancy still exists in their latent representations.
Transformer-based entropy models capture intricate spatial or channel correlations, but their reliance on self-attention mechanisms introduces high latency and computational overhead~\cite{liu2023learned, li2024frequency, jiang2023mlic, koyuncu2022contextformer}.

\IEEEpubidadjcol 

\begin{figure*}[t]
  \centering
   \includegraphics[width=.86\linewidth]{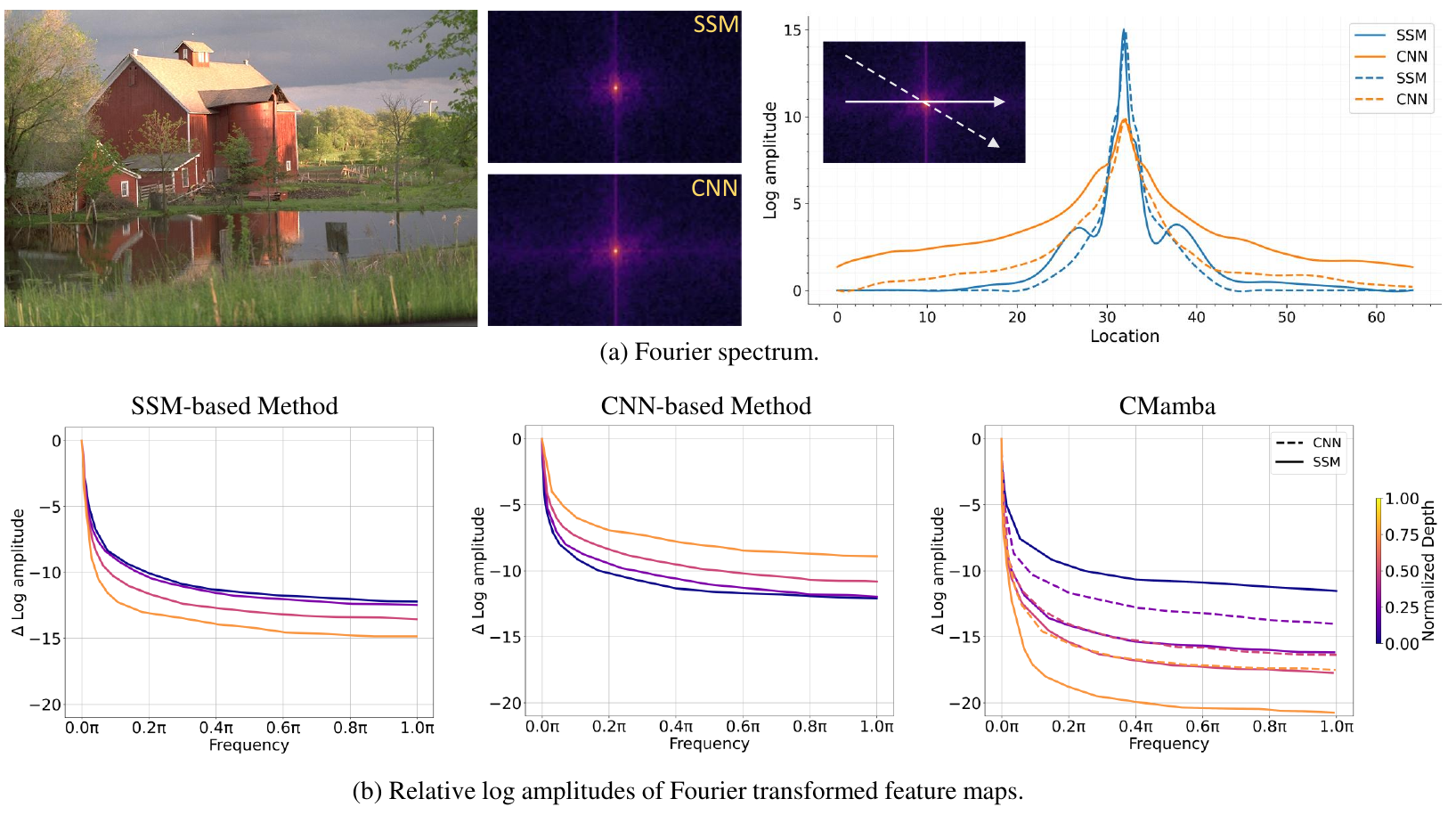}
   \vspace{-1em}
   \caption{
   The Fourier spectrum comparisons between SSMs and CNNs.
   (a) The Fourier spectrum of features obtained from the SSM-based method\protect\hyperlink{foot1}{\textsuperscript{1}} and the CNN-based method (ChARM)~\cite{minnen2020channel} in the last block of the analysis transform $g_a(\cdot)$.
   (b) Relative log amplitudes of Fourier transformed feature maps\protect\hyperlink{foot2}{\textsuperscript{2}}~for different methods. 
   $\Delta$ log amplitude values indicate the averaged output of each block in $g_a(\cdot)$ on the Kodak dataset.   
   }
   \vspace{-1em}
   \label{fig1}
\end{figure*}

State Space Models (SSMs) have recently demonstrated superior performance on various vision and language tasks~\cite{gu2023mamba, zhu2024vision, liu2024vmamba}.
Inspired by the advancements in SSMs, we propose a hybrid CNNs and SSMs based image compression framework, dubbed \textbf{\textit{CMamba}}, to achieve better rate-distortion performance and computational efficiency.
Our CMamba consists of two components: (1) a Content-Adaptive SSM (CA-SSM) module and (2) a Context-Aware Entropy (CAE) module.

Due to the linear computational complexity of SSMs, we intend to employ them to model global content while preserving global receptive fields~\cite{liu2024vmamba}.
However, we observed that SSMs excel in modeling overall content but tend to lose high-frequency details. 
This issue gets worse as network depths increase, as shown in Fig.~\ref{fig1}(b).
Hence, solely relying on SSMs would lead to inferior compression performance.
To tackle this issue, our CA-SSM module incorporates SSMs and CNNs to capture both global content and local details as CNNs can effectively capture fine-grained local details~\cite{park2021vision,zou2022devil,liu2023learned}.
As shown in Fig.~\ref{fig1}(a), the feature extracted by CNNs contains more high-frequency details compared to that captured by SSMs.
Thus, we integrate a simple yet effective CNN, as a complementary component to SSMs, in our CA-SSM module.

In the CA-SSM module, we employ a dynamic fusion block that can adaptively fuse SSM features (\emph{i.e.}, global content features) and CNN features (\emph{i.e.}, local features). 
The dynamic fusion block learns to determine whether sufficient image details or global content are encoded or decoded and then produces fusion weights for SSM and CNN features, respectively. 
In this fashion, the global content and local detail features are fully exploited in encoding and decoding.

Our CAE module is designed to jointly model spatial and channel dependencies, and thus enables precise and efficient entropy modeling of latent representations in bitstream compression.
To be specific, in the spatial dimension, our CAE module leverages SSMs to parameterize the distribution of spatial content via a learnable Gaussian model, as SSMs are good at capturing global content while performing in linear complexity. 
Along the channel dimension, the inter-channel relationships in latent representations are captured via an autoregressive manner.
Considering the nature of bitstream transmission, we process each channel sequentially and use the hidden states of previously processed channels as condition to further reduce inter-channel dependency. 
In this way, channel-wise prior knowledge can be exploited to reduce inter-channel redundancy, leading to lower bitrates in entropy coding.

\hypertarget{foot1}{\footnotetext[1]{
The convolutional layers in the main path~\cite{minnen2020channel} are replaced with visual state space blocks~\cite{liu2024vmamba}.
The models are optimized with Mean Squared Error (MSE), and $\lambda$ is set to 0.05.
}}

\hypertarget{foot2}{\footnotetext[2]{
The $\Delta$ log amplitude is defined as the difference between the log amplitude at a normalized frequency of 0.0$\pi$ (center) and 1.0$\pi$ (boundary). 
For better visualization, only the half-diagonal components of two-dimensional Fourier-transformed feature maps are shown.
}}

To demonstrate the effectiveness of CMamba, we conduct extensive experiments on widely-used image compression benchmarks, \ie, Kodak~\cite{franzen1999kodak}, Tecnick~\cite{asuni2014testimages}, and CLIC~\cite{theis2020clic}.
CMamba achieves superior rate-distortion performance, and outperforms Versatile Video Coding (VVC)~\cite{bross2020versatile} by 14.95\%, 18.83\%, and 13.89\% on these three benchmarks, respectively. 
In particular, compared to the state-of-the-art LIC method~\cite{jiang2023mlicpp}, CMamba reduces parameters by 51.8\%, FLOPs by 28.1\%, and decoding time by 71.4\% on the Kodak dataset. 
The main contributions can be summarized as follows:
\begin{itemize}
    \item 
    We propose a hybrid Convolution and State Space Models based image compression framework, termed CMamba, and achieve better rate-distortion performance with low computational complexity.

    \item 
     We propose a Content-Adaptive SSM (CA-SSM) module that dynamically fuses global content from SSMs and local details from CNNs in encoding and decoding stages.

    \item 
    We design a Context-Aware Entropy (CAE) module that explicitly models spatial and channel dependencies, enabling precise and efficient entropy modeling of latent representations for bitstream compression.

\end{itemize}

\section{Related Work}
\subsection{Image Compression}

Image compression is a vital field in digital image processing, aimed at improving image storage and transmission efficiency.
Classical lossy image compression standards, such as JPEG~\cite{wallace1991jpeg}, BPG~\cite{bellard2018bpg}, and VVC~\cite{bross2020versatile}, rely on handcrafted rules and have been widely adopted.
Recently, learned image compression has made significant progress and achieved promising performance~\cite{balle2017end, song2021variable, rhee2022lc, lee2022dpict, cui2021asymmetric, ma2022end, ali2023towards, fu2023learned}.
Ball\'e~\etal~\cite{balle2017end} propose a pioneering end-to-end optimized image compression model, which significantly improves compression performance by leveraging CNNs.
Cheng~\etal~\cite{cheng2020learned} incorporate attention mechanisms into their compression network, thus enhancing the encoding of complex regions.
Xie~\etal~\cite{xie2021enhanced} utilize invertible neural networks (INNs) to mitigate the issue of information loss and achieve better compression.
Yang~\etal~\cite{yang2024lossy} propose a novel transform-coding-based lossy compression scheme using diffusion models.
Zhu~\etal~\cite{zhu2022transformer} and Zou~\etal~\cite{zou2022devil} propose transformer based image compression networks and obtain superior compression effectiveness compared to CNNs.
Liu~\etal~\cite{liu2023learned} integrate transformers and CNNs to harness both non-local and local modeling capabilities, enhancing the overall performance of image compression. Concurrent with our work, Qin~\etal~\cite{qin2024mambavc} investigate a pure SSM network for image compression.

In addition, several studies have been proposed to explore various entropy models to improve image compression.
Inspired by side information in image codecs, hyperprior is introduced to capture spatial dependencies in latent representations~\cite{balle2018variational}.
Driven by autoregression of probabilistic generative models, Minnen~\etal~\cite{minnen2018joint} predict latent representations from a causal context model along with a hyperprior.
Due to the time-consuming process of spatial scanning in autoregressive models, Minnen~\etal~\cite{minnen2020channel} propose a channel-wise autoregressive model as an alternative while He~\etal~\cite{he2021checkerboard} develop a checkerboard context model for parallel computing. 
Following these works, various adaptations of these methods have also been developed~\cite{he2022elic, jiang2023mlic, koyuncu2024efficient}. 
However, it remains a challenge to jointly model spatial and channel dependencies in an efficient manner.

\subsection{State Space Models}

State Space Models (SSMs) have shown their effectiveness in capturing the dynamics and dependencies~\cite{gu2020hippo, gu2021combining, goel2022s}.
To reduce excessive computational and memory requirements in SSMs, Gu~\etal~\cite{gu2021efficiently} constrain their parameters into a diagonal structure.
Subsequently, structured state space models have been proposed, such as complex-diagonal structures~\cite{gu2022parameterization, gupta2022diagonal}, multiple-input multiple-output configurations~\cite{smith2022simplified}, combinations of diagonal and low-rank operations~\cite{hasani2022liquid}, and gated activation functions~\cite{mehta2023long}.
Among them, Mamba introduces selective scanning and a hardware speed-up algorithm to facilitate efficient training and inference~\cite{gu2023mamba}.
Vim~\cite{zhu2024vision} is the first SSM-based model, as a general vision backbone, to address the limitations of Mamba in modeling image sequences.
VMamba~\cite{liu2024vmamba} introduces a cross-scan module to traverse the spatial domain and transform any non-causal visual image into ordered patch sequences.
Huang~\etal~\cite{huang2024localmamba} propose a novel local scanning strategy that divides images into distinct windows to capture local and global dependencies.
Mamba has been explored for its potential in various vision tasks, including image restoration~\cite{guo2024mambair, cheng2024activating, deng2024cu, shi2024vmambair}, point cloud processing~\cite{li20243dmambacomplete, liang2024pointmamba, liu2024point, zhang2024point}, video modeling~\cite{chen2024video, li2024videomamba, zou2024rhythmmamba}, and medical image analysis~\cite{ma2024u, yue2024medmamba, ma2024semi}, but how to effectively apply Mamba in image compression remains unexplored.   
\section{Preliminaries}

\begin{figure*}[t]
  \centering
   \includegraphics[width=.8\linewidth]{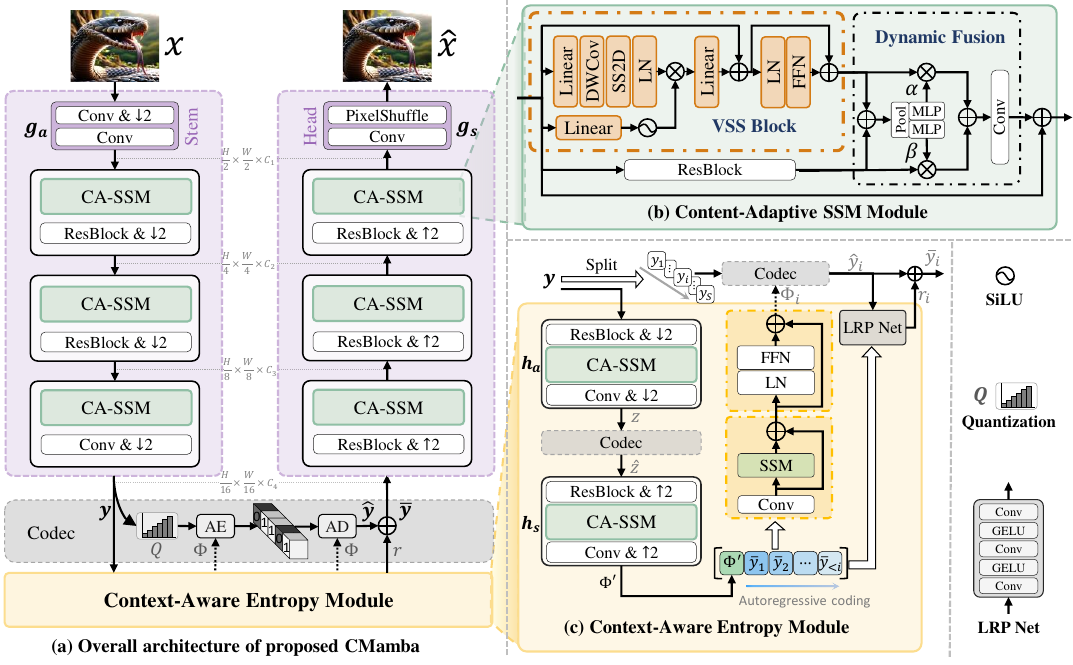}
   \caption{
   (a) 
   Overview of our proposed method. 
   (b) 
   Detailed design of our proposed Content-Adaptive SSM (CA-SSM) module.
   The CA-SSM module has two parallel paths (\ie, VSS block and ResBlock) to capture global content and local details, and then fuses these features dynamically.
   (c) 
   The detailed network architecture of our Context-Aware Entropy (CAE) module.
   The CAE module jointly models spatial and channel dependencies in latent representations $y$.
    }
    \vspace{-1em}
   \label{fig2}
\end{figure*}

\noindent \textbf{Learned Image Compression (LIC).~} 
Here, we provide a brief overview of LIC.
In general, LIC follows a three-stage paradigm: nonlinear transformation, quantization, and entropy coding.
The nonlinear transformation consists of an analysis transform and a synthesis transform.
The analysis transform $g_a(\cdot)$ maps an image $x$ into a latent representation $y$.
Then, quantization $Q(\cdot)$ converts the latent representation $y$ to its discrete form.
Since the quantization process introduces clipping errors in the latent representation $r = y - Q(y)$, it would lead to distortion in the reconstructed image. 
As suggested in~\cite{minnen2020channel}, the quantization error $r$ can be estimated via a latent residual prediction network.
Finally, the rectified latent representation $\bar{y}=\hat{y} + r$ is transformed back to a reconstructed image $\hat{x}$ using the synthesis transform $g_s(\cdot)$.
The process is summarized as follows:
\begin{equation}
y = g_a(x; \phi),\
\hat{y} = Q(y),\
\hat{x} = g_s(\hat{y}+r; \theta), 
\label{eq1}
\end{equation}
where $\phi$ and $\theta$ represent the optimized parameters for the analysis and synthesis transforms, respectively.

The latent representation $y$ is assumed to follow a Gaussian distribution, characterized by parameters $\Phi$, \ie, mean $\mu$ and standard deviation $\sigma$ (aka, scale).
In the channel-wise autoregressive entropy model, side information $z$ is introduced as an additional prior to estimate the probability distribution of the latent representation $y$~\cite{minnen2020channel}.
To be specific, a hyper-encoder $h_a(\cdot)$ takes the latent representation $y$ as input to generate the side information.
Then, $z$ will also be quantized as $\hat{z}$ via $Q(\cdot)$. Next, a hyper-prior decoder $h_s(\cdot)$ is applied to the quantized side information $\hat{z}$ to derive a hyper-prior $\Phi^{'}$.
This process is formulated as follows:
\begin{equation}
z = h_a(y; \phi_h),\
\hat{z} = Q(z),\
\Phi^{'} = h_s(\hat{z}; \theta_h).
\label{eq2}
\end{equation}
Subsequently, the latent representation $y$ is split into $S$ groups along the channel dimension, denoted as $\{y_1, ... , y_S\}$.
The hyper-prior $\Phi^{'}$ and decoded groups $\hat{y}_{s<i}$ are used to estimate parameters $\Phi_i$ of Gaussian distributions for the current group $\hat{y}_i$.
As a result, the Gaussian probability $p(\hat{y}_i|\Phi^{'}, \hat{y}_{s<i})$ is modeled in an autoregressive manner.

To train the overall learned image compression model, we adopt rate-distortion as the optimization objective, defined as:
\begin{align}
\mathcal{L} &= R(\hat{y}) + R(\hat{z}) + \lambda \cdot D(x, \hat{x}) \notag \\
            &= \mathbb{E} \left[ -\log_2 \left( p(\hat{y} | \hat{z}) \right) \right] + \notag \\
            &\quad \mathbb{E} \left[ -\log_2 \left( p(\hat{z}) \right) \right] + \lambda \cdot \mathbb{E}\left[ d(x, \hat{x}) \right], 
\label{eq3}
\end{align}
where $\lambda$ controls the trade-off between rate and distortion.
$R$ represents the bit rate of $\hat{y}$ and $\hat{z}$, and $d(x, \hat{x})$ is the distortion between the input image $x$ and reconstructed image $\hat{x}$.

\noindent \textbf{State Space Models (SSMs).~}
Continuous-time SSMs can be regarded as a Linear Time-Invariant (LTI) system that transforms a sequential input $x(t)\in \mathbb{R}$ to an output $y(t)\in \mathbb{R}$ via a hidden state $h(t)\in \mathbb{R}^N$. 
It is formulated as follows:
\begin{equation}
\begin{aligned}
h'(t) &= A h(t) + B x(t), \\
y(t) &= C h(t) + D x(t), 
\label{eq4}
\end{aligned}
\end{equation}
where $h'(t)$ denotes the first derivative of the hidden state $h(t)$ with respect to time $t$. 
$A\in \mathbb{R}^{N \times N}$, $B\in \mathbb{R}^{N \times 1}$, and $C\in \mathbb{R}^{1 \times N}$ are coefficient matrices for the LTI system. 
$D\in \mathbb{R}$ is a feedthrough parameter~\cite{hespanha2018linear}.

To be integrated into deep models, continuous-time SSMs need to be discretized. 
This process uses a times-cale parameter $\Delta$ for transforming the $A$ and $B$ into their discretized forms.
Consequently, Eqn.~\eqref{eq4} can be discretized via the zero-order hold (ZOH) as follows:
\begin{equation}
\begin{aligned}
h_k &= e^{\Delta A} h_{k-1} + (\Delta A)^{-1} (e^{\Delta A} - I) \cdot \Delta B x_k, \\
y_k &= C h_k + D x_k.
\end{aligned}
\label{eq5}
\end{equation}

\section{Methodology}
Our proposed hybrid Convolution and State Space Models (SSMs) based image compression framework is illustrated in Fig.~\ref{fig2}.
Specifically, we design two components, \ie, a Content-Adaptive SSM (CA-SSM) module (marked by the green blocks) and a Context-Aware Entropy (CAE) module (marked by the yellow block).
Our CA-SSM module (Sec.~\ref{sec CA-SSM}) is designed to dynamically fuse global content and local details extracted by SSMs and CNNs, respectively.
Then, our CAE module (Sec.~\ref{sec CAE}) is presented to model spatial and channel dependencies jointly.
These dependencies facilitate effective yet efficient entropy modeling of latent representations for bitstream compression.

\subsection{Content-Adaptive SSM Module}
\label{sec CA-SSM}

SSMs have demonstrated superior performance on various vision and language tasks~\cite{gu2023mamba, zhu2024vision, liu2024vmamba, guo2024mambair}, and they offer a global receptive field with linear complexity. 
Intuitively, SSMs could be a better candidate backbone for image compression as they have the potential to balance compression effectiveness and efficiency.
Hence, the Content-Adaptive SSM (CA-SSM) module is designed to fully exploit the linear computational complexity of State Space Models (SSMs) and their global content modeling capability for image compression.

Our CA-SSM incorporates a Visual State Space (VSS) block to capture global content.
The VSS block adopts a 2D-Selective-Scan (SS2D) layer to traverse the spatial domain and convert any non-causal visual image into ordered patch sequences~\cite{liu2024vmamba}.
This scanning strategy facilitates SSMs in handling visual data without compromising the field of reception.
The SS2D layer within the VSS block unfolds feature patches along four directions, producing four distinct sequences. 
Then, these sequences are processed via SSMs, and the output features from different directions are merged to reconstruct a complete feature map.
Given an input feature $\mathcal{F}_\textit{IN}$, the output feature $\mathcal{F}_\textit{OUT}$ of the VSS can be expressed as:
\vspace{-0.5em}
\begin{equation}
\vspace{-0.5em}
\begin{aligned}
\mathcal{F}_\textit{SS2D} &= \textit{LN}(f_\textit{ss2d}(\sigma (w_1(\textit{LN}(\mathcal{F}_\textit{IN}))))), \\
\mathcal{A} &= \sigma (w_2 \textit{LN}(\mathcal{F}_\textit{IN})), \\
\mathcal{F}_{1} &= w_3( \mathcal{F}_\textit{SS2D} \odot \mathcal{A}) + \mathcal{F}_\textit{IN}, \\
\mathcal{F}_\textit{OUT} &= w_4( \textit{LN} (\mathcal{F}_{1}) ) + \mathcal{F}_{1},
\end{aligned}
\label{eq6}
\end{equation}
where $w_1$, $w_2$, $w_3$, and $w_4$ are learned parameters, 
$\textit{LN}(\cdot)$ denotes layer normalization,
$\sigma(\cdot)$ represents the \textit{SiLU} activation function~\cite{ramachandran2017searching}, 
and $\odot$ denotes the element-wise product. 
The function $f_\textit{ss2d}(\cdot)$ refers to an SS2D operation, defined as:
\vspace{-0.5em}
\begin{equation}
\vspace{-0.5em}
\begin{aligned}
x_{v} &= f_\textit{exp}(x_{in}, v), \\
\bar{x}_v &= f_\textit{ssm}(x_{v}), \\
x_{out} &= f_\textit{mrg}(\bar{x}_v \mid v \in V),
\end{aligned}
\label{eq7}
\end{equation}
where $V=\{1, 2, 3, 4\}$ represents a set of four different scanning directions, and $v \in V$ denotes a specific scanning direction.
Here, $f_\textit{exp}(\cdot)$ performs the scan expansion in direction $v$. Then, the output $x_v$ of $f_\textit{exp}(\cdot)$ is passed to SSMs, and $\bar{x}_v$ is estimated by the function $f_\textit{ssm}(\cdot)$, defined in Eqn.~\eqref{eq5}.
$f_\textit{mrg}(\cdot)$ combines the outputs in all the directions~\cite{liu2024vmamba}.

Although SSMs effectively model the overall content, they often struggle to preserve high-frequency image details, as illustrated in Fig.~\ref{fig1}(a). 
Moreover, as network depths increase, this issue would get worse, as shown in Fig.~\ref{fig1}(b).
As a result, solely relying on SSMs would lead to inferior compression performance.
To tackle this issue, we propose to integrate a CNN block in our CA-SSM module as CNNs excel at capturing fine-grained local details~\cite{park2021vision, zou2022devil, liu2023learned}.
As illustrated in Fig.~\ref{fig1}(a), features extracted by CNNs contain more high-frequency details compared to those from SSMs.
Therefore, a simple yet effective ResBlock~\cite{he2016deep} is adopted to capture local details.
While a VSS block models the global content of an image, the ResBlock plays a complementary role to the VSS block in our CA-SSM module. 
In doing so, an input feature $x\in \mathbb{R}^{C \times H \times W}$ is processed through parallel branches of SSMs and CNNs, producing features $\mathcal{F}_\textit{SSM}$ and $\mathcal{F}_\textit{CNN}$, as shown in Fig.~\ref{fig2}(b).

Moreover, we employ a dynamic fusion block to fuse SSM features (\ie, global content features) and CNN features (\ie, local features) in our CA-SSM module. 
It learns to determine which features are more beneficial in improving rate-distortion performance.
In this way, our CA-SSM module seamlessly integrates global content features and local detail features in encoding and decoding.
Specifically, we first merge $\mathcal{F}_\textit{SSM}$ and $\mathcal{F}_\textit{CNN}$, and then apply a global max pooling operation to derive channel-wise representations, denoted by $\mathcal{F}_\textit{S} = f_{gp}(\mathcal{F}_\textit{SSM} + \mathcal{F}_\textit{CNN})$. 
Subsequently, $\mathcal{F}_\textit{S}$ is processed via a multilayer perceptron and a softmax operation to obtain corresponding attention weights $\alpha$ and $\beta$. 
Finally, these attention weights are used to modulate the features extracted from SSMs and CNNs dynamically.
Thus, the output $y$ of our CA-SSM module can be expressed as:
\vspace{-0.5em}
\begin{equation}
\begin{aligned}
y &= w(\alpha \cdot \mathcal{F}_\textit{SSM} + \beta \cdot \mathcal{F}_\textit{CNN}), \\
\alpha &= \frac{\exp(\mathcal{F}_\alpha)}{\exp(\mathcal{F}_\alpha) + \exp(\mathcal{F}_\beta)}, \\
\beta &= \frac{\exp(\mathcal{F}_\beta)}{\exp(\mathcal{F}_\alpha) + \exp(\mathcal{F}_\beta)}, \\
\mathcal{F}_\alpha  & = w_{mlp_1}(\mathcal{F}_\textit{S}), \quad \mathcal{F}_\beta = w_{mlp_2}(\mathcal{F}_\textit{S}), 
\end{aligned}
\label{eq8}
\end{equation}
where $w \in \mathbb{R} ^{C\times C}$ is a learnable parameter, $w_{mlp_1}$ and $w_{mlp_2}$ are the weights of the multilayer perceptions.

\subsection{Context-Aware Entropy Module}
\label{sec CAE}

As shown in Fig.~\ref{fig2}(c), CAE is designed to address the following challenges in the entropy model:
(1) how to precisely model content distribution while minimizing the bit number, and
(2) how to enhance the efficiency of entropy coding.
We design the CAE module to jointly model spatial and channel dependencies, thus facilitating precise and efficient entropy modeling of latent representations.

In the spatial dimension, our CAE leverages SSMs to parameterize the spatial content via Gaussian modeling due to its linear complexity in modeling global content dependencies.
Moreover, hardware speed-up algorithms are adopted in SSMs, including selective scan, kernel fusion, and recomputation, to aid efficient training and inference~\cite{gu2023mamba, zhu2024vision, liu2024vmamba, li2024videomamba}. 
Considering the sequential decoding nature of bitstreams, the inter-channel relations within latent representations are modeled autoregressively. 
In this way, the efficiency of encoding and decoding will not be significantly delayed.
To be specific, each channel is processed sequentially and conditioned on the prior derived from previously processed channels.
In this way, the channel-wise prior knowledge can be exploited to reduce inter-channel redundancy, thus minimizing bitrates.

Given a latent representation $y$, we first split it into $S$ groups along the channel dimension, \ie, $\{y_1, ... , y_S\}$.
To compress $y_{i}$, we concatenate the hyper-prior $\Phi^{'}$ (Eqn.~\eqref{eq2}) with the previous decoded groups $\bar{y}_{s<i}$.
These concatenated features are then processed via SSMs to estimate the Gaussian distribution parameters $\Phi_{i}$.
$\Phi_{i}$ is used to determine the Cumulative Distribution Function (CDF) for arithmetic coding.
Accurate estimation of $\Phi_{i}$ can reduce entropy and thus decrease the bit number for compression.
This process is defined as follows:
\begin{equation}
\begin{aligned}
\mathcal{F}_\textit{SQ} &= w_{sq}([\Phi^{'}, \bar{y}_{<i}]), \\
\mathcal{F}_\textit{SSM} &= f_\textit{ssm}(\mathcal{F}_\textit{SQ}) + \mathcal{F}_\textit{SQ}, \\
\Phi_{i} &= w_\textit{ffn}(\textit{LN}(\mathcal{F}_\textit{SSM})) + \mathcal{F}_\textit{SSM}, 
\end{aligned}
\label{eq9}
\end{equation}
where $w_{sq}$ is a learnable parameter,
and $\left[\cdot\right]$ indicates the concatenation operation.
The $w_\textit{ffn}$ is a learnable parameter of a Feed-Forward Network (FFN).
Next, a Latent Residual Prediction (LRP) network is employed to reduce this quantization error.
The error $r$ introduced by the quantization operation is defined as $r = y - Q(y)$.
The LRP network predicts $r$ using the hyper-prior $\Phi^{'}$ and previously decoded groups (\ie, $\bar{y}_{s<i}$ and $\hat{y}_{i}$).  
\section{Experiments}

\subsection{Experimental Setup}

\noindent \textbf{Training.~} 
Following the previous work~\cite{zou2022devil}, we train the proposed CMamba model on the OpenImages dataset~\cite{krasin2017openimages}. 
Our CMamba is trained for 50 epochs using the Adam optimizer~\cite{kingma2014adam}.
Each batch contains 8 patches with the size of $256 \times 256$ randomly cropped from the training images.
The learning rate is initialized as $1e^{-4}$. 
After 40 epochs, the learning rate is reduced to $1e^{-5}$ for 5 epochs. 
Finally, we train the model for the last 5 epochs with a larger crop size of $512 \times 512$, maintaining the learning rate at $1e^{-5}$.

Our model is optimized by the rate-distortion loss as illustrated in Eqn.~\eqref{eq3}.
The distortion $D$ is quantified by two quality metrics, \ie, mean square error (MSE) and multi-scale structural similarity index (MS-SSIM)\footnote[3]{Here, we represent the MS-SSIM by $-10\log_{10}\left( 1-\textit{MS-SSIM} \right)$ for a clearer comparison.}.
The Lagrangian multipliers used for training MSE-optimized models are $\left\{25, 35, 67, 130, 250, 500 \right\}\times1e^{-4}$, and those for MS-SSIM-optimized models are $\left\{3, 5, 8, 16, 36, 64 \right\}$.

\noindent \textbf{Evaluation.~}
We evaluate our model on three benchmark datasets, \ie, Kodak dataset~\cite{franzen1999kodak} with the image size of $768 \times 512$, Tecnick testset~\cite{asuni2014testimages} with the image size of $1200 \times 1200$, and CLIC Professional Validation dataset~\cite{theis2020clic} with 2K resolution. 
PSNR and MS-SSIM are used to evaluate the quality of reconstructed images, and bits per pixel (bpp) is used to evaluate Bitrate.
Besides rate-distortion curves, we also evaluate different models using BD-Rate~\cite{tan2015video}, which describes the average Bitrate savings for the same reconstruction quality.
All experiments are conducted on an NVIDIA GeForce RTX 3090 Ti and an Intel i9-12900.

\subsection{Rate-Distortion Performance}

\begin{figure*}[t]
  \centering
   \includegraphics[width=0.95\linewidth]{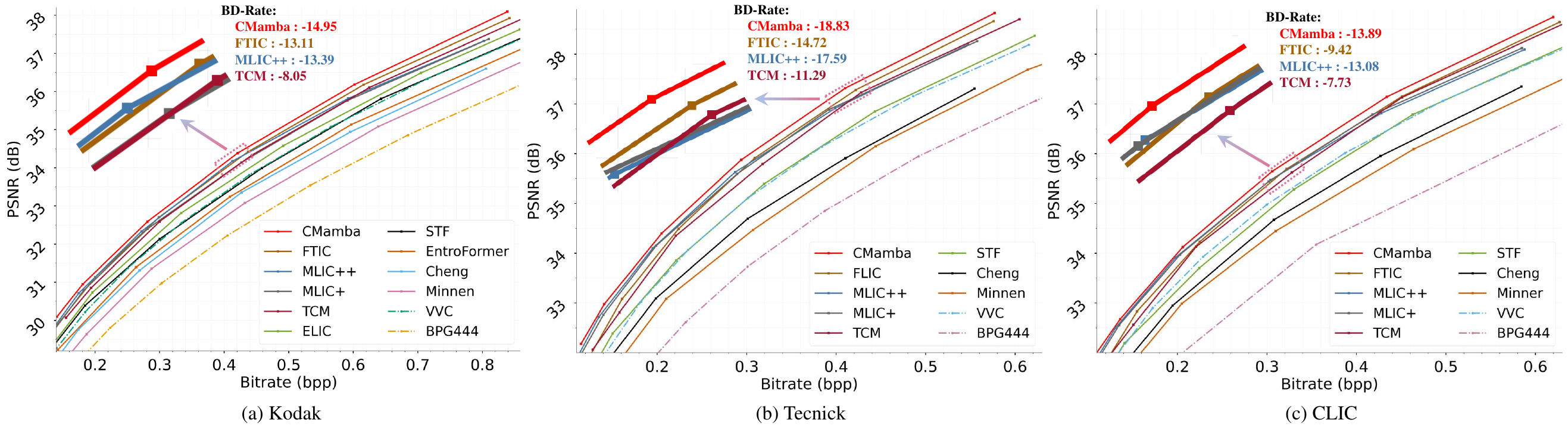}
   \vspace{-1em}
   \caption{
   PSNR-Bitrate curves evaluated on Kodak, Tecnick, and CLIC datasets.
   The compared methods include state-of-the-art LIC models and handcrafted codecs.
   LIC models are optimized with MSE.
   }
   \vspace{-1em}
   \label{fig3}
\end{figure*}

\begin{table*}[t]
\centering
\setlength{\tabcolsep}{8pt}
\renewcommand{\arraystretch}{1.05}
\caption{
Rate-distortion performance and coding complexity are evaluated on the Kodak, Tecnick, and CLIC datasets.
\textbf{Enc.} and \textbf{Dec.} denote inference latency for encoding and decoding respectively.
\textbf{Tot.} represents the total inference latency.
The \textbf{BD-Rate} is presented for rate-distortion performance comparison with VVC as the anchor. 
$\downarrow$ indicates that a lower value is better.
}
\begin{tabular}{clcccccc}
\toprule
\multicolumn{1}{c}{\multirow{2}{*}{\textbf{Dataset}}}  & \multicolumn{1}{c}{\multirow{2}{*}{\textbf{Method}}}  & \multicolumn{3}{c}{\textbf{Latency(ms)} $\downarrow$}& \multirow{2}{*}{\textbf{\#Params(/M)} $\downarrow$}& \multirow{2}{*}{\textbf{Flops} $\downarrow$} & \multirow{2}{*}{\textbf{BD-Rate(\%)} $\downarrow$} \\ \cmidrule(l{.5em}r{.5em}){3-5}
  & & \textbf{Enc.}& \textbf{Dec.} & \textbf{Tot.}\\  \midrule \midrule    
\multicolumn{1}{c}{\multirow{11}{*}{\textbf{Kodak}}}  & Minnen~\cite{minnen2018joint} \textcolor{gray}{\textit{NeurIPS'18}}& \textgreater{} 1000 &  \textgreater{} 1000& \textgreater{} 1000 & 20.15 &  176.79G &  \cellcolor[HTML]{EFEFEF} +15.15\\
\multicolumn{1}{c}{}  & Cheng~\cite{cheng2020learned} \textcolor{gray}{\textit{CVPR'20}}& \textgreater{} 1000 &   \textgreater{} 1000&  \textgreater{} 1000 &  27.55  &   403.27G & \cellcolor[HTML]{EFEFEF} +7.94\\
\multicolumn{1}{c}{}  & EntroFormer~\cite{qian2021entroformer} \textcolor{gray}{\textit{ICLR'22}}& \textgreater{} 1000&   \textgreater{} 1000&   \textgreater{} 1000 &  45.00&  - &  \cellcolor[HTML]{EFEFEF} +4.73\\
\multicolumn{1}{c}{}  & STF~\cite{zou2022devil} \textcolor{gray}{\textit{CVPR'22}}   & 72&    68&   140 &  99.86     &  200.11G & \cellcolor[HTML]{EFEFEF} -2.48\\
\multicolumn{1}{c}{}  & ELIC~\cite{he2022elic} \textcolor{gray}{\textit{CVPR'22}}   & 71&    92&   163 &  36.90&  327.12G &  \cellcolor[HTML]{EFEFEF} -5.95\\
\multicolumn{1}{c}{}  & TCM~\cite{liu2023learned} \textcolor{gray}{\textit{CVPR'23}}   & 108&   112&   220 &  76.57&  700.65G &  \cellcolor[HTML]{EFEFEF} -8.05\\
\multicolumn{1}{c}{}  & MLIC+~\cite{jiang2023mlic} \textcolor{gray}{\textit{MM'23}}      &     -&   -&    -&    -&  -&  \cellcolor[HTML]{EFEFEF} -11.39\\
\multicolumn{1}{c}{}  & FTIC~\cite{li2024frequency} \textcolor{gray}{\textit{ICLR'24}}   &  99&  110&  209 &    70.97&   490.00G &  \cellcolor[HTML]{EFEFEF} -13.11\\
\multicolumn{1}{c}{}  & MLIC++~\cite{jiang2023mlicpp} \textcolor{gray}{\textit{NCW'23}}  &  164 &  182&  346&  116.70&  494.18G &  \cellcolor[HTML]{EFEFEF} -13.39 \\
\multicolumn{1}{c}{}  & CMamba (Ours)  & 95&  52&  147 &   56.21 &  355.31G &   \cellcolor[HTML]{EFEFEF} \textbf{-14.95}  \\ \cmidrule{2-8} 
\multicolumn{1}{c}{}  & VVC   & \textgreater{} 1000     &140   &\textgreater{} 1000     & -    & - & \cellcolor[HTML]{EFEFEF} 0     \\ \midrule \midrule

\multicolumn{1}{c}{\multirow{9}{*}{\textbf{Tecnick}}}  & Minnen~\cite{minnen2018joint} \textcolor{gray}{\textit{NeurIPS'18}}& \textgreater{} 1000 & \textgreater{} 1000 & \textgreater{} 1000 & 20.15 & 664.80G & \cellcolor[HTML]{EFEFEF} +15.01 \\
\multicolumn{1}{c}{}  & Cheng~\cite{cheng2020learned} \textcolor{gray}{\textit{CVPR'20}}& \textgreater{} 1000 & \textgreater{} 1000 & \textgreater{} 1000 &  27.55  &  1.52T  & \cellcolor[HTML]{EFEFEF} +8.82 \\
\multicolumn{1}{c}{}  & STF~\cite{zou2022devil} \textcolor{gray}{\textit{CVPR'22}}   & 226 & 197 & 423 &  99.86     & 752.50G & \cellcolor[HTML]{EFEFEF} -2.14 \\
\multicolumn{1}{c}{}  & TCM~\cite{liu2023learned} \textcolor{gray}{\textit{CVPR'23}} & 389 & 364 & 753 &  76.57& 2.92T & \cellcolor[HTML]{EFEFEF} -11.29 \\
\multicolumn{1}{c}{}  & MLIC+~\cite{jiang2023mlic} \textcolor{gray}{\textit{MM'23}}      & - & - & - & - & - & \cellcolor[HTML]{EFEFEF} -16.38 \\
\multicolumn{1}{c}{}  & FTIC~\cite{li2024frequency} \textcolor{gray}{\textit{ICLR'24}}   & \textgreater{} 1000 & \textgreater{} 1000 & \textgreater{} 1000 &    70.97&  -  & \cellcolor[HTML]{EFEFEF} -14.72 \\
\multicolumn{1}{c}{}  & MLIC++~\cite{jiang2023mlicpp} \textcolor{gray}{\textit{NCW'23}}  & 372 & 398 & 770 &  116.70&  1.86T & \cellcolor[HTML]{EFEFEF} -17.59 \\
\multicolumn{1}{c}{}  & CMamba (Ours)  & 353 & 134 & 487 &   56.21 & 1.34T & \cellcolor[HTML]{EFEFEF} \textbf{-18.83}  \\ \cmidrule{2-8}
\multicolumn{1}{c}{}  & VVC   & \textgreater{} 1000 & 222 & \textgreater{} 1000 & -    & - & \cellcolor[HTML]{EFEFEF} 0     \\ \midrule \midrule

\multicolumn{1}{c}{\multirow{9}{*}{\textbf{CLIC}}}  & Minnen~\cite{minnen2018joint} \textcolor{gray}{\textit{NeurIPS'18}}& \textgreater{} 1000 &  \textgreater{} 1000 & \textgreater{} 1000 & 20.15 & 1.04T & \cellcolor[HTML]{EFEFEF} +16.90 \\
\multicolumn{1}{c}{}  & Cheng~\cite{cheng2020learned} \textcolor{gray}{\textit{CVPR'20}}& \textgreater{} 1000 & \textgreater{} 1000 & \textgreater{} 1000 &  27.55  &  2.38T  & \cellcolor[HTML]{EFEFEF} +11.63 \\
\multicolumn{1}{c}{}  & STF~\cite{zou2022devil} \textcolor{gray}{\textit{CVPR'22}}   & 294 &  227 & 521 &  99.86     & 1.18T & \cellcolor[HTML]{EFEFEF} +0.56 \\
\multicolumn{1}{c}{}  & TCM~\cite{liu2023learned} \textcolor{gray}{\textit{CVPR'23}} & 567 &  540 & \textgreater{} 1000 &  76.57& 4.23T & \cellcolor[HTML]{EFEFEF} -7.73 \\
\multicolumn{1}{c}{}  & MLIC+~\cite{jiang2023mlic} \textcolor{gray}{\textit{MM'23}}      & - & - & - & - & - & \cellcolor[HTML]{EFEFEF} -12.56 \\
\multicolumn{1}{c}{}  & FTIC~\cite{li2024frequency} \textcolor{gray}{\textit{ICLR'24}}   & \textgreater{} 1000 & \textgreater{} 1000 & \textgreater{} 1000 &    70.97&  -  & \cellcolor[HTML]{EFEFEF} -9.42 \\
\multicolumn{1}{c}{}  & MLIC++~\cite{jiang2023mlicpp} \textcolor{gray}{\textit{NCW'23}}  & 521 & 548 & \textgreater{} 1000 &  116.70& 2.92T & \cellcolor[HTML]{EFEFEF} -13.08 \\
\multicolumn{1}{c}{}  & CMamba (Ours)  & 503 & 191 & 694 & 56.21 & 2.09T &  \cellcolor[HTML]{EFEFEF} \textbf{-13.89} \\ \cmidrule{2-8}
\multicolumn{1}{c}{}  & VVC   & \textgreater{} 1000 & 254 & \textgreater{} 1000 & -    & - & \cellcolor[HTML]{EFEFEF} 0     \\ 
\bottomrule
\end{tabular}
\label{tab1}
\end{table*}

We compare our method with state-of-the-art (SoTA) image compression algorithms, including traditional image codecs Better Portable Graphics (BPG)~\cite{bellard2018bpg} and Versatile Video Coding (VVC) intra (VTM 17.0)~\cite{bross2020versatile}, as well as LIC models~\cite{minnen2018joint, cheng2020learned, qian2021entroformer, he2022elic, zou2022devil, jiang2023mlic, liu2023learned, li2024frequency, jiang2023mlicpp}.

Fig.~\ref{fig3} and Table~\ref{tab1} present the MSE optimized rate-distortion performance on Kodak, Tecnick, and CLIC datasets. Fig.~\ref{fig5} demonstrates the performance optimized by MS-SSIM on the Kodak dataset.
These results demonstrate that our method outperforms prior methods across all three datasets. 
To get quantitative results, we present the BD-Rate~\cite{tan2015video} computed from PSNR-Bitrate curves as the quantitative metric.
The anchor rate-distortion performance is set as the benchmark achieved by Versatile Video Coding (VVC) intra (VTM 17.0)~\cite{bross2020versatile} on different datasets (BD-Rate = 0\%). 
Our method achieves improvements of 14.95\%, 18.83\%, and 13.89\% in BD-Rate compared to VVC on Kodak, Tecnick, and CLIC datasets, respectively.
We also provide the BD-Rate for several SoTA image compression methods in Fig.~\ref{fig3} and Fig.~\ref{fig5}.
As seen in these figures, our CMamba outperforms other SoTA methods in rate-distortion performance.

Furthermore, we conduct comparative experiments to validate the efficiency of the proposed CMamba across multiple metrics, including latency, parameters, and FLOPs.
As shown in Table~\ref{tab1}, our method demonstrates substantial improvements on the Kodak dataset, achieving 51.8\% reduction in parameters, 28.1\% decrease in FLOPs, and 71.4\% reduction in decoding time compared to the SoTA LIC method~\cite{jiang2023mlicpp}.
Overall, our CMamba attains superior rate-distortion performance and significantly reduces computational complexity compared to the state-of-the-art.

\begin{figure*}[t]
  \centering
   \includegraphics[width=.85\linewidth]{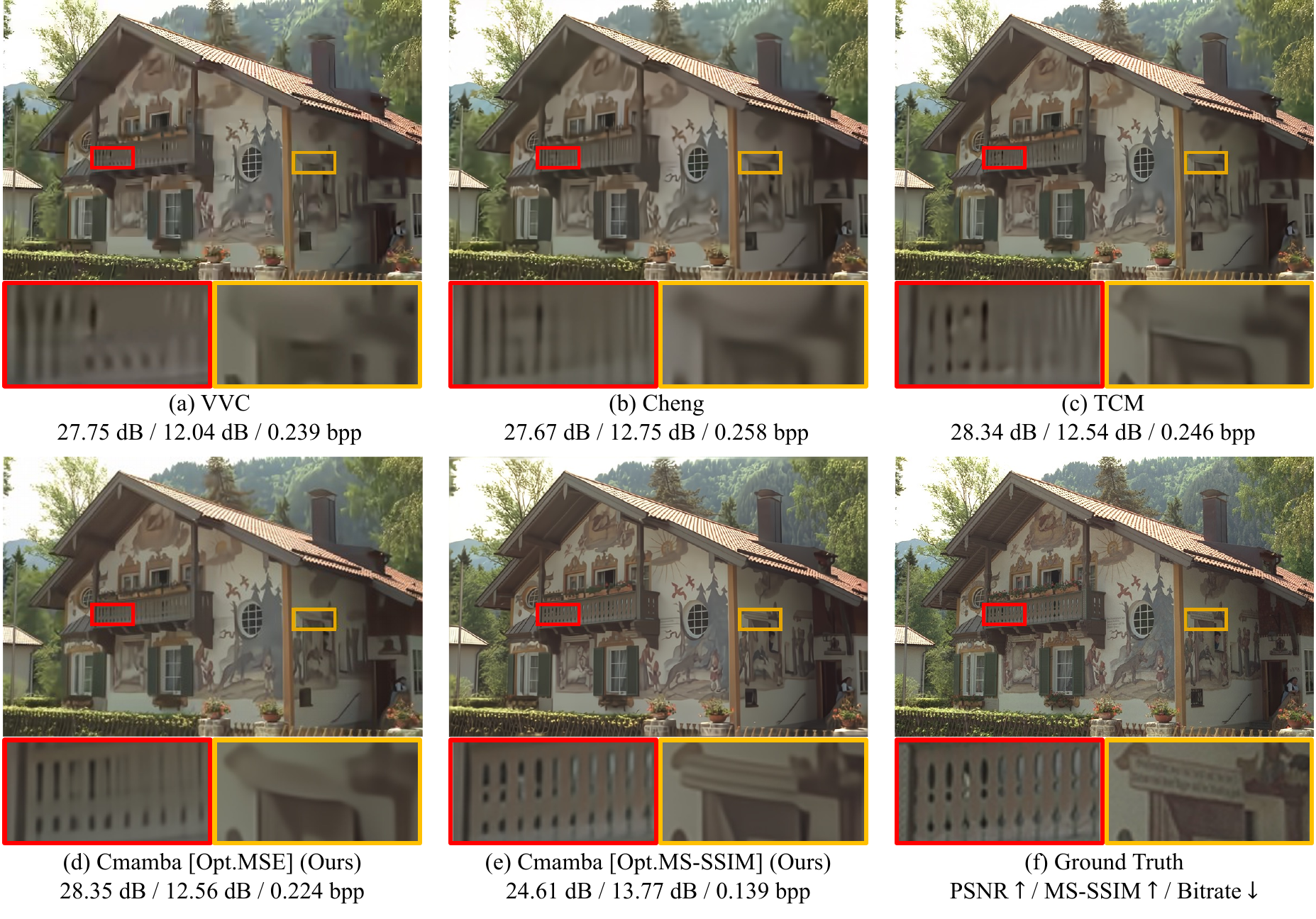}
   \vspace{-1em}
   \caption{
    Visual comparison of the decompressed \textit{kodim24.png} image from the Kodak dataset using various compression methods. 
    Opt.MSE and Opt.MS-SSIM indicate that a model is optimized with MSE and MS-SSIM, respectively. 
    More visual comparisons are provided in the supplementary materials.
   }
   \label{fig4}
\end{figure*}

\begin{figure}[t]
  \centering
   \includegraphics[width=.75\linewidth]{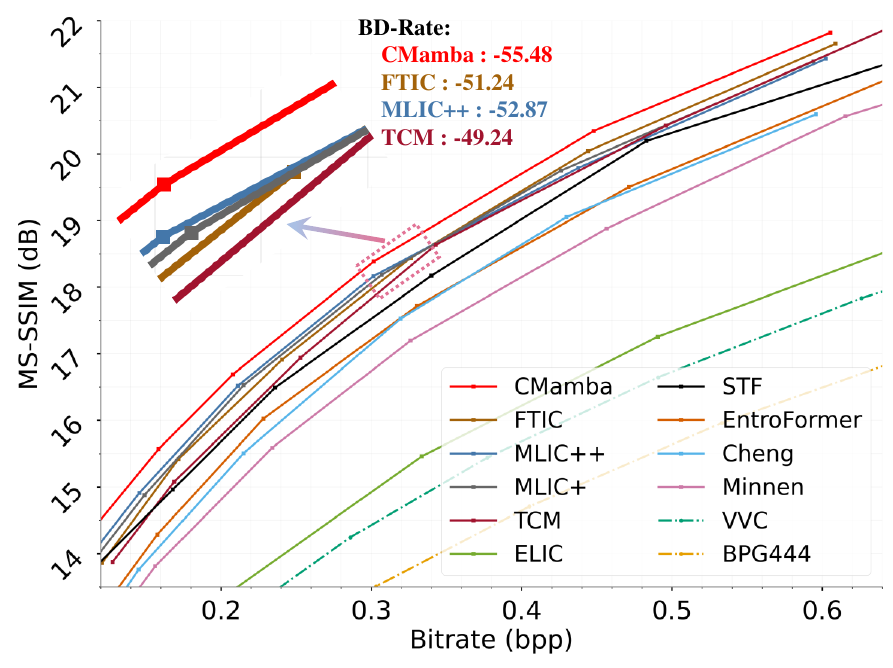}
   \vspace{-1em}
   \caption{
   Rate-distortion performance evaluated on the Kodak dataset.
   All the models are optimized with MS-SSIM.
   }
   \vspace{-1em}
   \label{fig5}
\end{figure}

\subsection{Qualitative Results}

To demonstrate that our method can produce visually appealing results, we provide visualizations of decompressed images for a qualitative comparison in Fig.~\ref{fig4}. 
The PSNR, MS-SSIM, and Bitrate values are indicated along with each sub-image label for additional quantitative reference.
Compared to TCM~\cite{liu2023learned}, CMamba~[Opt.MSE] preserves more details with a smaller Bitrate, such as sharper textures of the balcony railing (\textcolor[RGB]{244, 0, 0}{red box}) and mural details (\textcolor[RGB]{223, 165, 0}{yellow box}).
In the corresponding quantitative results, CMamba~[Opt.MSE] achieves a PSNR of 28.35 dB, an MS-SSIM of 12.56 dB, and a bitrate of 0.224 bpp, outperforming TCM, which achieves a PSNR of 28.34 dB, an MS-SSIM of 12.54 dB, and a bitrate of 0.246 bpp, respectively.
More importantly, the CMamba~[Opt.MS-SSIM] achieves better visual quality with a lower Bitrate (0.139 bpp) compared to other methods.

\subsection{Ablation Studies}

\begin{table}[t]
\centering
\footnotesize
\caption{
Ablation studies of the CA-SSM and CAE modules are evaluated on the Kodak dataset. 
The baseline configuration includes only the VSS Block and ChARM.
}
\renewcommand{\arraystretch}{1.05}
\begin{tabular}{c|cccc|c}
\toprule
 \textbf{CA-SSM} & & \checkmark& &\checkmark &\multirow{2}{*}{\textit{VVC}} \\
 \textbf{CAE}  &  &  &  \checkmark& \checkmark& \\ \midrule \midrule
 \textbf{Enc.(/ms)} $\downarrow$&      90&    94&   92&     95& \textgreater{} 1000 \\
 \textbf{Dec.(/ms)} $\downarrow$&      48&    50&   48&     52& 140                \\
 \textbf{Tot.(/ms)} $\downarrow$&     138&    144&  140&     147& \textgreater{} 1000 \\
 \textbf{\#Params(/M)} $\downarrow$&65.17&  64.33&  57.60&   56.21& - \\
 \textbf{FLOPs(/G)} $\downarrow$&  453.20& 367,76&  440.82&  355.29& - \\
 \textbf{BD-Rate(\%)} $\downarrow$& \cellcolor[HTML]{EFEFEF} -6.97& \cellcolor[HTML]{EFEFEF} -12.91 & \cellcolor[HTML]{EFEFEF} -10.83& \cellcolor[HTML]{EFEFEF} \textbf{-14.95} & \cellcolor[HTML]{EFEFEF} 0    \\
\bottomrule
\end{tabular}
\label{tab2}
\end{table}

We conduct ablation studies to demonstrate the effectiveness of our CA-SSM and CAE modules. 
Specifically, we replace the CA-SSM module and the CAE module with the VSS block~\cite{liu2024vmamba} and ChARM~\cite{minnen2020channel} to serve as the baseline model. 
As shown in Table~\ref{tab2}, the proposed CA-SSM module significantly improves the rate-distortion performance, saving 12.91\% BD-Rate, while maintaining low encoding (94 ms) and decoding (50 ms) time by dynamically integrating the advantages of SSMs and CNNs. 
Furthermore, the CAE module further improves the rate-distortion performance to -14.95\% BD-Rate with fewer parameters (56.21M) and fewer computational costs (355.29G FLOPs) compared to ChARM.
This implies that the combination of CA-SSM and CAE not only achieves superior rate-distortion performance but also attains efficiency in terms of computational complexity and inference speed.
In addition, we further analyze the contributions of each component in our CA-SSM and CAE modules.

\subsubsection{Analysis of the CA-SSM Module Design}

To further verify the design of the CA-SSM module, we conduct experiments with other architectures (\ie, CNN, Swin, SSM, and Swin \& CNN) and fusion methods (\ie, Summation and Concatenation), as presented in Table~\ref{tab3}.
In our experimental configuration, \textit{CNN}, \textit{Swin}, and \textit{SSM} denote that the CA-SSM module is replaced with the corresponding layer, respectively, while maintaining approximately the same number of parameters.
The \textit{Swin \& CNN} indicates that the VSS block within the CA-SSM module is substituted with the Swin Transformer block~\cite{liu2021swin}. 
For fusion methods, \textbf{\textit{Sum}} and \textbf{\textit{Concat}} refer to configurations where features are fused via summation or concatenation operations, rather than dynamic fusion.
All configurations utilize ChARM~\cite{minnen2020channel} as the entropy module.
The comparison demonstrates that our CA-SSM module outperforms all alternatives, achieving the best performance with a 12.91\% BD-Rate saving and 64.33M parameters.

\begin{table}[t]
\centering
\setlength{\tabcolsep}{3pt}
\caption{
Comparative analysis of different backbones and fusion methods in the content-adaptive SSM (CA-SSM) module on the Kodak dataset.
}
\renewcommand{\arraystretch}{1.05}
\begin{tabular}{cc|cc}
\toprule
\multicolumn{2}{c|}{\textbf{Method}} & \textbf{\#Params(/M)} $\downarrow$ & \textbf{BD-Rate(\%)} $\downarrow$ \\ \midrule \midrule
\multicolumn{1}{c|}{\multirow{5}{*}{\textbf{Backbone}}}   & CNN &  65.75 \scriptsize{\textcolor[HTML]{6c757d}{+1.42}}& -7.17 \scriptsize{\textcolor[HTML]{6c757d}{+5.74}} \\
\multicolumn{1}{c|}{}                          & Swin &  65.12 \scriptsize{\textcolor[HTML]{6c757d}{+0.79}}& -7.84 \scriptsize{\textcolor[HTML]{6c757d}{+5.07}} \\ 
\multicolumn{1}{c|}{}                          & SSM &  65.17 \scriptsize{\textcolor[HTML]{6c757d}{+0.84}}& -9.18 \scriptsize{\textcolor[HTML]{6c757d}{+3.73}} \\ 
\multicolumn{1}{c|}{} & Swin \& CNN &  70.34 \scriptsize{\textcolor[HTML]{6c757d}{+6.01}} &  -10.52 \scriptsize{\textcolor[HTML]{6c757d}{+2.39}}  \\
\multicolumn{1}{c|}{}                          & SSM \& CNN (Ours) & \textbf{64.33} &  \textbf{-12.91} \\ \midrule
\multicolumn{1}{c|}{\multirow{3}{*}{\textbf{Fusion}}} & Sum &     68.83 \scriptsize{\textcolor[HTML]{6c757d}{+4.50}} &  -12.33 \scriptsize{\textcolor[HTML]{6c757d}{+0.58}}  \\
\multicolumn{1}{c|}{}                               & Concat &    74.26 \scriptsize{\textcolor[HTML]{6c757d}{+9.93}} &  -12.65 \scriptsize{\textcolor[HTML]{6c757d}{+0.26}}  \\ 
\multicolumn{1}{c|}{}                               & Dynamic Fusion (Ours) &  \textbf{64.33} &  \textbf{-12.91} \\ \midrule
\multicolumn{2}{c|}{VVC} &           -    &    0        \\ \bottomrule
\end{tabular}
\label{tab3}
\end{table}

\begin{table}[t]
\centering
\setlength{\tabcolsep}{5pt}
\caption{
Comparison of proposed context-aware entropy (CAE) module against various entropy models on the Kodak dataset.
}
\renewcommand{\arraystretch}{1.05}
\begin{tabular}{l|ccc}
\toprule
\multicolumn{1}{c|}{\textbf{Method}} & \textbf{\#Params(/M)} $\downarrow$ & \textbf{Latency(ms)} $\downarrow$ & \textbf{BD-Rate(\%)} $\downarrow$ \\ \midrule \midrule 
\noalign{\vskip -0.5mm}
\hspace{-2mm}Ours $g_a$ and $g_s$ &         &          \\
\noalign{\vskip -0.5mm}
\hspace{1mm}+ ChARM~\cite{minnen2020channel}  & 64.33 \scriptsize{\textcolor[HTML]{6c757d}{+8.12}}  & 144 \scriptsize{\textcolor[HTML]{6c757d}{-3}} &  -12.91 \scriptsize{\textcolor[HTML]{6c757d}{+2.04}}  \\
\hspace{1mm}+ ELIC~\cite{he2022elic}          & {53.41} \scriptsize{\textcolor[HTML]{6c757d}{-2.80}} & 158 \scriptsize{\textcolor[HTML]{6c757d}{+11}} &  -13.08 \scriptsize{\textcolor[HTML]{6c757d}{+1.87}}  \\
\hspace{1mm}+ T-CA~\cite{li2024frequency}     & 77.67 \scriptsize{\textcolor[HTML]{6c757d}{+21.46}}  & 204 \scriptsize{\textcolor[HTML]{6c757d}{+57}} &  -13.84 \scriptsize{\textcolor[HTML]{6c757d}{+1.11}}  \\
\hspace{1mm}+ TCM~\cite{liu2023learned}       & 87.24 \scriptsize{\textcolor[HTML]{6c757d}{+31.03}}  & 212 \scriptsize{\textcolor[HTML]{6c757d}{+65}} &  \underline{-14.19} \scriptsize{\textcolor[HTML]{6c757d}{+0.76}}   \\
\hspace{1mm}+ CAE (Ours)    & {56.21}  & 147 &  \textbf{-14.95} \\ \midrule
\multicolumn{1}{c|}{VVC} & -        & \textgreater{} 1000 &  0        \\ \bottomrule
\end{tabular}
\label{tab4}
\end{table}

\begin{figure}[t]
  \centering
   \includegraphics[width=.95\linewidth]{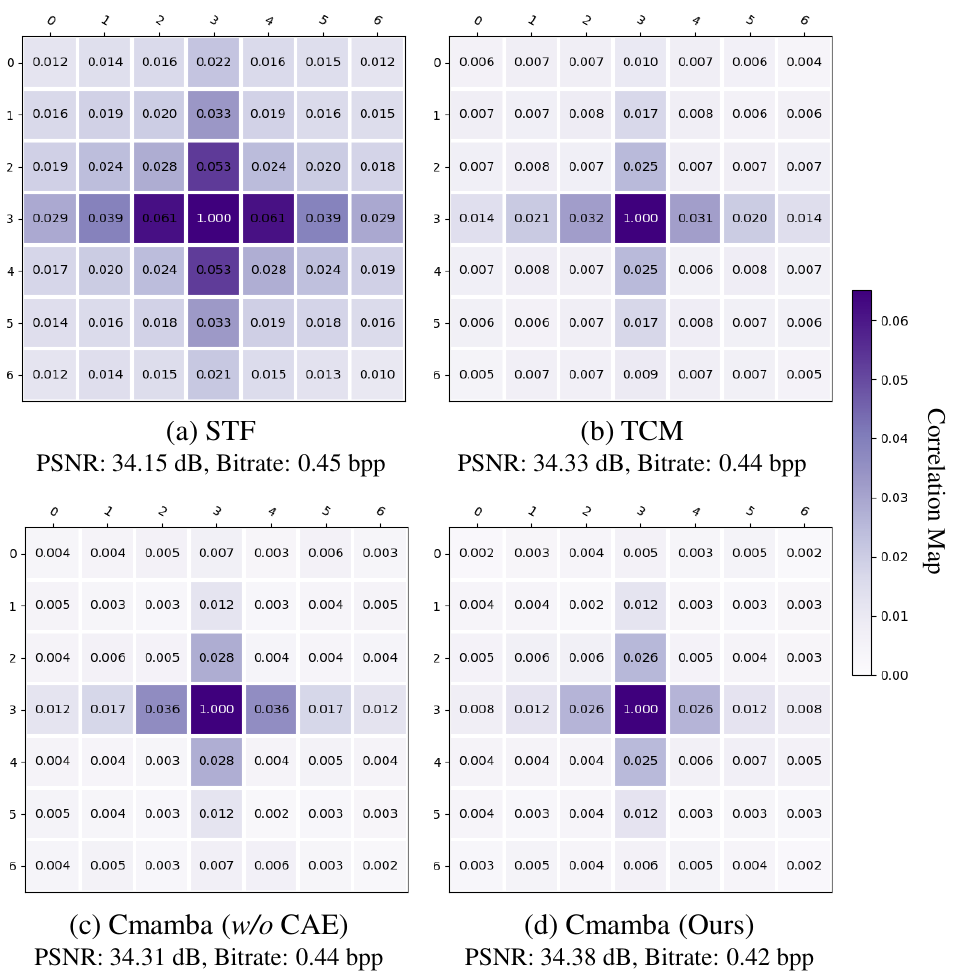}
   \vspace{-1em}
   \caption{
   The spatial correlation map of $(y-\mu)/\sigma$ with models trained at $\lambda=0.013$. 
   The value with index $(i, j)$ corresponds to the normalized cross-correlation of latent representation at spatial locations $(w, h)$ and $(w + i, h + j)$, averaged across all latent elements of all images on the Kodak dataset.
   $w/o$ denotes the substitution of the CAE module with ChARM.
   }
   \vspace{-1em}
   \label{fig7}
\end{figure}

\subsubsection{Analysis of the CAE Module Design}

To demonstrate the superiority of our CAE module in entropy modeling, we conduct experiments with other entropy models~\cite{minnen2020channel, he2022elic, liu2023learned, li2024frequency}, as shown in Table~\ref{tab4}. 
The CAE module harnesses an SSM-enhanced hyperprior and group-wise conditioning to enhance compression efficiency and reduce redundancy. 
In Table~\ref{tab4}, the CAE module achieves superior rate-distortion performance and much fewer parameters compared to the second-best entropy model, \ie, TCM~\cite{liu2023learned}.
This experiment indicates that the CAE module not only outperforms existing entropy models in terms of rate-distortion performance but also improves compression effectiveness.

\begin{table}[t]
\centering
\setlength{\tabcolsep}{3pt}
\caption{
Ablation studies of the proposed context-aware entropy (CAE) module on the Kodak dataset.
\textbf{S} denotes spatial dependencies.
\textbf{C} represents channel dependencies.
\textbf{CAR} indicates channel-wise autoregressive modeling.
}
\renewcommand{\arraystretch}{1.05}
\begin{tabular}{cc|ccc}
\toprule
\multicolumn{2}{c|}{\textbf{Method}} & \textbf{\#Params(/M)} $\downarrow$  & \textbf{Latency(ms)} $\downarrow$& \textbf{BD-Rate(\%)} $\downarrow$ \\ \midrule \midrule
\multicolumn{1}{c|}{\multirow{3}{*}{S}}   & CNN  & 72.07 & 135 & -13.02 \\ 
\multicolumn{1}{c|}{}                     & Swin & 72.87 & 191 & -14.49 \\ 
\multicolumn{1}{c|}{}                     & SSM (Ours)  & 56.21 & 147 &  \textbf{-14.95} \\ \midrule
\multicolumn{1}{c|}{\multirow{2}{*}{C}} & \textit{w/o} CAR & 71.24 & 108 & +1.05 \\
\multicolumn{1}{c|}{}                   & \textit{w} CAR (Ours)   &  56.21& 147 &  \textbf{-14.95} \\ \midrule
\multicolumn{2}{c|}{VVC} &   -  &  \textgreater{} 1000 & 0 \\ \bottomrule
\end{tabular}
\label{tab5}
\end{table}

Furthermore, we conduct experiments to carefully verify the efficacy of the CAE module, as presented in Table~\ref{tab5}. 
In particular, we compare different approaches, including CNNs, Swin Transformers, and SSMs, to capture spatial dependencies.
Meanwhile, we also evaluate the effectiveness of channel dependencies.
The channel dependencies are captured in an autoregressive manner.
\textit{w/o} CAR means to directly estimate the distribution parameters of latent representation $y$ via a Mean \& Scale Hyperprior~\cite{minnen2018joint}.
This experiment highlights that the CAE module achieves significant improvements in compression performance by jointly modeling spatial and channel dependencies while maintaining efficiency.

In addition, our CAE module estimates the mean $\mu$ and scale $\sigma$ of latent representation $y$ via a hyperprior to eliminate the redundancy of latent representation $y$~\cite{balle2018variational, cheng2020learned}.
Therefore, we conduct the following analysis for latent correlation.
The latent correlation reflects the redundancy in $(y - \mu) / \sigma$. 
The spatial correlation maps in Fig.~\ref{fig7} illustrate the capabilities of different models in redundancy reduction. 
STF~(Fig.~\ref{fig7}(a)) and TCM~(Fig.~\ref{fig7}(b)) show higher correlations indicating less effective redundancy removal. 
In contrast, CMamba (\textit{w/o} CAE)~(Fig.~\ref{fig7}(c)) demonstrates improved redundancy reduction. 
Notably, our CMamba~(Fig.~\ref{fig7}(d)) achieves the lowest correlation across spatial positions benefiting from its global Effective Receptive Field and the integration of the CAE module. 
These results confirm the superiority of CMamba in decorrelating latent representations, thus leading to better compression performance with a lower Bitrate (0.42 bpp) and higher PSNR (34.38 dB).
  
\section{Conclusion}

In this paper, we introduced CMamba, a hybrid image compression framework that combines the strengths of Convolutional Neural Networks (CNNs) and State Space Models (SSMs) to achieve a balance between high rate-distortion performance and low computational complexity. 
The proposed Content-Adaptive SSM (CA-SSM) module effectively integrates global content from SSMs with local details from CNNs, ensuring the preservation of critical image features during compression. 
Additionally, the Context-Aware Entropy (CAE) module enhances spatial and channel compression efficiency by reducing redundancies in latent representations, leveraging SSMs for spatial parameterization and an autoregressive approach for channel redundancy reduction. 
Notably, CMamba achieved substantial reductions in parameters, FLOPs, and decoding time, reinforcing its practical applicability in scenarios requiring efficient and high-performance image compression. 
By advancing the integration of SSMs and CNNs via the CA-SSM and CAE modules, CMamba represents a meaningful step forward in the field of learned image compression.

\bibliographystyle{IEEEtran}
\bibliography{IEEEabrv, main}

\vfill

\end{document}